\documentstyle[aps]{revtex}

\begin{document}

\title{The ground state of the lithium atom in strong 
magnetic fields}
\author{M. V. Ivanov\dag\  and P. Schmelcher}
\address{Theoretische Chemie, Physikalisch--Chemisches Institut,
Universit\"at Heidelberg, INF 253, D-69120 Heidelberg, 
Federal Republic of Germany\\
\dag Permanent address: Institute of Precambrian Geology and Geochronology,
Russian Academy of Sciences,
Nab. Makarova 2, St. Petersburg 199034, Russia
}

\date{\today}
\maketitle

\begin{abstract}
The ground and some excited states of the Li atom 
in external uniform magnetic fields are calculated 
by means of our 2D mesh Hartree-Fock method for field strengths 
ranging from zero up to $2.35\cdot 10^8$T. 
With increasing field strength the ground state undergoes two
transitions involving three different electronic configurations:
for weak fields the ground state configuration arises from the
field-free $1s^22s$ configuration, for intermediate fields from the  
$1s^22p_{-1}$ configuration and in high fields the $1s2p_{-1}3d_{-2}$ 
electronic configuration is responsible for the properties of the atom.
The transition field strengths are determined.
Calculations on the ground state of the $\rm Li^+$ ion 
allow us to describe the field-dependent ionization energy of the Li atom.
Some general arguments on the ground states of multi-electron atoms in strong
magnetic fields are provided.
\end{abstract}


\section{Introduction}

The behaviour and properties of atoms in strong magnetic fields
is a subject of increasing interest. On the o.h.s. this is 
motivated by the astrophysical discovery of strong fields
on white dwarfs and neutron stars \cite{NStar1,NStar2,Whdwarf1}
and on the o.h.s. the competition of the diamagnetic and Coulombic
interaction causes a rich variety of complex properties which are
of interest on their own.
Investigations on the electronic structure in the presence of 
a magnetic field appear to be quite complicated 
due to the mixed geometry of this quantum problem 
(mixing of spherical and cylindrical symmetry).
There are many works on the hydrogen atom 
(for a list of references see \cite{Fri89,RWHR,Ivanov88,Kra96})
and several works on the He atom as well as He-like ions 
\cite{Ivanov91,TKBHRW,Ivanov94,JonesOrtiz,JonesOrtiz97}.
Other atoms however have been investigated only in a very few cases 
\cite{JonesOrtiz,Ivanov97,Neuhauser}.

For the hydrogen atom the impact of the mixed symmetry is 
particularly evident and at the same time pronounced 
in the intermediate field regime
for which the magnetic and Coulomb forces are comparable.
For different electronic degrees of excitation of the atom the intermediate
regime is met for different absolute values of the field strength.
For the ground state the boundaries of this regime 
can be defined in a rough manner
as the range 
$\gamma=~0.2-20$  ($\gamma =B/B_0$, $B$ 
is the magnetic field strength, 
$B_0=\hbar c/ea_0^2=2.3505 {\cdot} 10^5$T; 
atomic units will be used in the following). 
With increasing degree of excitation the domain of the intermediate fields 
lowers correspondingly and becomes, as a rule, 
wider on a logarithmic scale of $\gamma$. 
Both early \cite{Garstang} and more recent works 
\cite{RWHR,SimVir78,Friedrich,Fonte,Schmidt}
on the hydrogen atom have used different approaches for relatively 
weak fields (the Coulomb force prevails over the magnetic force) 
and for very strong fields where the Coulomb force can be 
considered as weak in comparison with the magnetic forces
(adiabatic limit).  
In early works 
the Coulomb field was considered in this limit actually 
as perturbation for a free electron 
in a superstrong magnetic field. 
The motion of an electron parallel to the magnetic field is 
governed in the adiabatic approximation \cite{ElLoudon} by a 1D quasi-Coulomb 
potential with a parameter, dependent on the magnetic field strength. 
The detailed calculations of the hydrogen energy levels carried out 
by R\"osner {\it et al} \cite{RWHR} also retain the separation 
of the magnetic field strength domains due to decomposing 
the electronic wave function in terms of either spherical 
(for weak fields) or cylindrical harmonics 
(for strong fields). 
A powerful method to obtain comprehensive results on low-lying energy levels 
in the intermediate regime in particular for the hydrogen atom is provided by
mesh methods \cite{Ivanov88}.

For atoms with several electrons the problem of the mixed symmetries 
is even more intricate than for hydrogen because different electrons feel 
very different Coulomb forces, i.e. possess different single-particle energies,
and the domain of the intermediate fields therefore 
appears to be the sum of the
intermediate domains for the separate electrons. 

There exist several investigations on two-electron atoms in the literature
\cite{Ivanov91,TKBHRW,Ivanov94,JonesOrtiz,JonesOrtiz97,Neuhauser,VinBay89,Larsen,Virtamo76,Gadiyak,Mueller75}. 
The majority of them deals with the adiabatic limit in superstrong fields.
Most of the early works are Hartree-Fock (HF) calculations for the strong 
field domain. 
There are also several variational calculations for the low-field domain 
\cite{Larsen,Henry74,Surmelian74} 
including  calculations by Larsen \cite{Larsen} made at 
$\gamma\leq 2$ for He atom 
and at $\gamma\leq 5$ for $\rm H^-$. 
The latter calculations can be used for evaluations of the 
correlation energy in 
the low-field domain. 
HF calculations \cite{TKBHRW} are carried out 
analogously to 
the approach in ref.\cite{RWHR} with applying 
two different sets of basis functions 
to the high- and low-field domains. As a result 
of the complicated geometry in the
intermediate regime this approach inherently suffers from very slow convergence
properties with respect to the energy eigenvalues and
yields therefore only a very low accuracy. 
Accurate calculations for arbitrary field strengths were carried out in refs.
\cite{Ivanov91,Ivanov94} by the 2D mesh HF method. Investigations on the
ground state as well as a number of excited states of helium including
the correlation energy have very recently been performed via a Quantum Monte
Carlo approach \cite{JonesOrtiz97}. Very recently benchmark results with
a precision of $10^{-4}-10^{-6}$ for the energy levels
have been obtained for a large number of
excited states with different symmetries using a configuration
interaction approach with an anisotropic Gaussian basis set \cite{Bec98}.

For the lithium atom which is the subject of the present work 
there exists only one recent investigation 
by Jones {\it et al} \cite{JonesOrtiz}.
It contains calculations for the ground state and 
a few low-lying states of the Li 
atom at weak and intermediate fields. 
Precise Hartree-Fock results for several states 
in weak fields and quite satisfactory results for the  intermediate 
region are presented in this work. However their
basis functions did not allow to perform calculations for stronger fields. 
An attempt to define the sequence 
of the electronic ground state configurations 
which are different for different regimes of the field strength has also
been undertaken in this work.
However a detailed qualitative analysis of the high-field ground state 
configuration was not carried out. 
As a result the high-field ground state electronic configuration 
and the transition point to this configuration from the intermediate one 
is still an open question.

In the current work we apply a fully numerical 2D Hartree-Fock method 
to the problem of the Li atom in magnetic fields of arbitrary strength.
This method enables us performing calculations for various states and with 
approximately equal precision for weak, intermediate and 
superstrong magnetic fields. 
Our main focus is the ground state of the Li atom and its ionization energies. 
To this end several electronic configurations of the Li atom and two
configurations of the ${\rm Li^+}$ ion are studied.

\section{Formulation of the problem and method of solution}

We solve the electronic Schr\"odinger equation for the lithium atom in 
a magnetic field under the assumption of an infinitely heavy nucleus
in the (unrestricted) Hartree-Fock approximation. 
The solution is established in the cylindrical coordinate system
$(\rho,\phi,z)$ with the $z$-axis oriented along the magnetic field.
We prescribe to each electron a definite value of the magnetic 
quantum number $m_\mu$. 
Each single-electron wave function $\Psi_\mu$ depends on the variables
$\phi$ and $(\rho,z)$ 
\begin{eqnarray}
\Psi_\mu(\rho,\phi,z)=(2\pi)^{-1/2}e^{-i m_\mu\phi}\psi_\mu(z,\rho)
\label{eq:phiout}
\end{eqnarray}
where $\mu=1,2,3$ is the numbering of the electrons. 
The resulting partial differential equations for $\psi_\mu(z,\rho)$ 
and the formulae for the Coulomb and exchange potentials 
have been presented in 
ref.\cite{Ivanov94}.

The one-particle equations for the wave functions $\psi_\mu(z,\rho)$ 
are solved 
by means of the fully numerical mesh method described in refs.
\cite{Ivanov88,Ivanov94}. 
The new feature which distinguishes the present calculations from 
those described in ref.\cite{Ivanov94} is the method of calculation 
of the Coulomb and exchange integrals. In the present work as well as in 
ref.\cite{Ivanov97} 
we obtain these potentials as solutions 
of the corresponding Poisson equations. 
The problem of the boundary conditions for the Poisson equation
as well as 
the problem of simultaneously solving 
Poissons equations on the same meshes with 
Schr\"odinger-like equations for the wave functions $\psi_\mu(z,\rho)$
have been discussed in ref.\cite{Ivanov94}.
In the present approach these problems are solved 
by using special forms of non-uniform meshes. 
Solutions to the Poisson equation on separate meshes contain 
some errors $\delta_P$ associated with an inaccurate description of the 
potential far from the nucleus. 
However due to the special form of the function $\delta_P(h)$ 
for these meshes 
(where $h$ is a formal mesh step) 
the errors do not show up in the final results for the energy 
and other physical quantities, which we obtain by means of the Richardson 
extrapolation procedure (polynomial extrapolation to $h=0$ 
\cite{Ivanov88,ZhVychMat}).
An additional improvement with respect to the precision of our numerical 
calculations 
of the integrals is achieved by solving the Poisson equation not for 
the whole charge distribution but for the total distribution minus 
some properly chosen charge distribution 
with known analytical solution to the Poisson equation. 
Both of these approaches will be described in detail in a separate work.

Our mesh approach is flexible enough to yield precise 
results for arbitrary field strengths.
Some minor decrease of the precision 
appears in very strong magnetic fields. 
This phenomenon 
is due to a growing difference in the 
binding energies ${\epsilon_B}_\mu$ 
of single-electron wave functions belonging to the same 
electronic configuration 
\begin{eqnarray}
{\epsilon_B}_\mu=(m_\mu+|m_\mu|+2s_{z\mu}+1)\gamma/2-\epsilon_\mu
\label{eq:ebinone}
\end{eqnarray}
where $\epsilon_\mu$ is the single-electron energy 
and $s_{z\mu}$ is the spin $z$-projection. 
This results in big differences with respect 
to the spatial extension of the density 
distribution for different electrons. 
This difference is important for the configurations $1s^22s$,
$1s2s2p_{-1}$ and 
$1s2p_0 2p_{-1}$ and  
is not important for 
$1s2p_{-1}3d_{-2}$ 
and $1s^2 2p_{-1}$
because all the single-electron energies for the latter states 
are of the same order of magnitude. 
The precision of our results depends, of course, on the number of mesh nodes 
and can be improved in calculations with denser meshes. 
The most dense meshes which we could use in the present calculations 
had $120 \times 120$ nodes. These meshes were used for the states 
$1s^22s$, $1s 2s 2p_{-1}$ and 
$1s 2p_0 2p_{-1}$ 
at fields $\gamma=500$ and $\gamma=1000$. 
For other states and weaker magnetic fields Richardson's sequences of 
meshes with maximal number $80\times 80$ or $60\times 60$ were 
sufficient.

\section{Ground state electronic configurations}

We start this section with a qualitative 
consideration of the problem of the atomic multi-electron ground states
in the limit of strong magnetic fields. 
It is clear that the state $1s^22s$ of the lithium atom is the 
ground state only for relatively weak fields.
The set of single-electron wave functions for constructing 
the HF ground state for the opposite case of extremely strong 
magnetic fields can be determined as follows.
The nuclear attraction energies and HF potentials 
(which determine the motion along $z$ axis)
are then small compared to the interaction energies with the magnetic field 
(which determines the motion perpendicular to the magnetic field 
and is responsible for the Landau zonal structure of the spectrum).
Thus, all the single-electron wave functions must correspond to the 
lowest Landau zones, i.e. $m_\mu\leq 0$
for all the electrons, and the system must be fully spin-polarized,
i.e. $s_{z\mu}= -{1\over2}$ ($\downarrow$). 
For the Coulomb central field the single-electron levels form
quasi 1D Coulomb series with the binding energy 
$E_B={1\over{2n_z^2}}$ for $n_z>0$ and
$E_B\rightarrow \infty$ for $n_z=0$,
where $n_z$ is the number of nodal
surfaces of the wave function, which cross the $z$ axis.
These relations between single-electron energies and the geometry 
of single-electron wave functions along with analogous relations for 
the field-free atom provide the basis for the 
following considerations.

It is evident, that the wave functions with $n_z=0$ have to be
choosen for the ground state at $\gamma \rightarrow \infty$.
Thus, for  $\gamma \rightarrow \infty$ the ground state of the Li atom
must be 
$1s\downarrow2p_{-1}\downarrow3d_{-2}\downarrow$. 
This state was not considered in \cite{JonesOrtiz} but only the
$1s\downarrow 2p_0\downarrow 2p_{-1}\downarrow$ 
configuration was presented. 
Analogously, the very high-field ground state for the C atom 
considered in \cite{JonesOrtiz} must be the state belonging 
to the configuration
$1s\downarrow2p_{-1}\downarrow3d_{-2}\downarrow4f_{-3}\downarrow5g_{-4}
\downarrow6h_{-5}\downarrow$.

The problem of the configuration of the 
ground state for the intermediate field region cannot be
solved without doing explicite calculations. 
Calculations in ref.\cite{JonesOrtiz} were carried out for 
configurations with the maximal single-electron principal quantum 
number $n\leq 2$. 
Under this restriction calculations for the states 
$1s^22s$,
$1s^2 2p_{-1}$, 
$1s\downarrow 2s\downarrow 2p_{-1}\downarrow$, and
$1s\downarrow 2p_0\downarrow 2p_{-1}\downarrow$ 
are sufficient to determine the set of intermediate ground states. 
Indeed, $1s^22s$ is the zero-field ground state. 
$1s^2 2p_{-1}$ is the lowest excited state 
of the field free atom and (contrary to $1s^22s$) all the 
single-electron wave functions of this state must have 
infinite binding energies in the infinite strong magnetic field. 
Moreover, this state has the largest binding energy $E_B$
\begin{eqnarray}
E_B={\sum_{\mu=1}^3(m_\mu+|m_\mu|+2s_{z\mu}+1)\gamma/2-E}
\label{eq:ebin}
\end{eqnarray}
in the strong field limit 
due to the fact that 
$\epsilon_B(1s)>\epsilon_B(2p_{-1})>\epsilon_B(3d_{-2})>\ldots$
in strong fields. 
(For $\gamma=1000$ one can obtain binding energies from 
table \ref{tab:ebgamma} as 
$E_B(1s^2 2p_{-1})=69.1569$ and $E_B(1s2p_{-1}3d_{-2})=60.0589$).  
The reader should note that the $1s^2 2p_{-1}$ 
configuration cannot represent the ground state in very strong fields
since it is not fully spin polarized. The state 
$1s\downarrow 2s\downarrow 2p_{-1}\downarrow$
is the lowest fully spin-polarized state with 
the single-electron principal quantum numbers $n_\mu \leq 2$ 
in weak fields and, at last, 
the state 
$1s\downarrow 2p_0\downarrow 2p_{-1}\downarrow$ 
which lies higher at $\gamma=0$ must 
become lower than 
$1s\downarrow 2s\downarrow 2p_{-1}\downarrow$
with increasing field strength.

Our calculations include the high-field ground state 
$1s\downarrow2p_{-1}\downarrow3d_{-2}\downarrow$ 
which contains one electron with $n=3$. 
In principle, also other configurations could be
considered as possible ground states for intermediate field strength.
Such configurations are $1s^23s$, $1s^23p_{-1}$, $1s^23d_{-2}$, 
$1s2s3s$, $1s2s3p_{-1}$, $1s2s3d_{-2}$, $1s2p_{-1}3s$, and $1s2p_{-1}3p_{-1}$.
Calculations for all these states are possible by means 
of our mesh HF method. However they are extremely tedious and
time consuming and have not been accomplished in the present work.
Indeed we will argue in the following that none of these 
states can be the ground state of the Li atom for intermediate field strength.

It is quite evident that for the configurations containing a
$1s^2$ pair of electrons the $1s^23s$ lies higher in energy than the $1s^22s$ 
configuration and that the $1s^23p_{-1}$ and $1s^23d_{-2}$ configuration
possess higher energy than the $1s^22p_{-1}$ configuration.
Thus, the states with $1s^2$ pairs can be excluded 
from our argumentation of the
ground state. Among the fully spin polarized configurations the levels of the
configurations $1s2p_{-1}3s$, $1s2s3p_{-1}$, $1s2s3d_{-2}$, 
and $1s2p_{-1}3p_{-1}$ 
are higher than that of the  $1s2s2p_{-1}$
configuration (two components of the configurations are identical
with those of $1s2s2p_{-1}$ and the third one is significantly higher). 
Thus from simple geometrical reasons only the $1s2s3s$ configuration
(mixed with the $1s2s3d_0$ configuration)
is {\it a priori} not excluded from becoming the intermediate ground state.
In weak magnetic fields this state lies slightly lower than 
other doubly excited and autoionizing states and 
in this regime it is the lowest
fully spin-polarized state. 
But the change of the ground state to the fully spin-polarized configuration 
takes place in the vicinity of $\gamma=2$ for which the $3s$ wave functions 
is much weaker bound  than the $3d_{-2}$, $2p_{-1}$ and even $2p_0$ orbitals. 
Due to this fact also the $1s2s3s$ configuration can be excluded
from becoming the ground state for any field strength. Indeed
our calculations show that this state becomes higher in energy than the
$1s2s2p_{-1}$ at $\gamma\approx 0.16$.

Thus, the set $1s^2 2p_{-1}$, 
$1s\downarrow 2s\downarrow 2p_{-1}\downarrow$, and
$1s\downarrow 2p_0\downarrow 2p_{-1}\downarrow$ 
along with weak-  $1s^22s$ and strong-field 
$1s\downarrow2p_{-1}\downarrow3d_{-2}\downarrow$ 
ground states is comprehensive for the determination 
of the ground state of the Li atom in a magnetic field of
arbitrary strength.

\section{Numerical Results}

The only work on the Li atom in a magnetic field with which we can
compare our results is ref.\cite{JonesOrtiz}. 
In this reference HF calculations were performed for weak and intermediate 
magnetic field strengths. Table \ref{tab:ebgamma}
contains the total energies obtained for the Li atom within
our calculations in comparison with the data obtained in \cite{JonesOrtiz}.
Our energy values coincide with those of  
ref.\cite{JonesOrtiz} for weak fields 
and lie substantially lower in the intermediate regime. 
At the upper boundary of the field region investigated in 
\cite{JonesOrtiz} the difference between \cite{JonesOrtiz} and 
our energies is 0.0239 for the $1s^22s$ state, 0.0205 for 
the $1s^22p_{-1}$ state, 
0.0870 for the $1s2s2p_{-1}$ state, and 0.0458 for the $1s2p_02p_{-1}$ state.

Our results on the total energies are illustrated in 
figures 1 and 2. 
These figures show in particular 
the ground state configurations
for the different regimes of the field strength.
One can conclude from table \ref{tab:ebgamma} and figures 1 and 2 
that the $1s^22s$ configuration represents the ground state for 
$0\leq\gamma<0.17633$, for $0.17633<\gamma<2.153$ 
the ground state configuration is 
$1s^2 2p_{-1}$, and for $\gamma>2.153$ 
the ground state configuration is 
$1s\downarrow2p_{-1}\downarrow3d_{-2}\downarrow$. 
The state $1s\downarrow 2p_0\downarrow 2p_{-1}\downarrow$ 
presented in \cite{JonesOrtiz} as the high field ground state 
appears not to be the ground state of the Li atom 
for any magnetic field strength.

Figure 3 presents spatial distributions of the total electronic densities 
for the ground state configurations of the lithium atom. 
In each row these densities are presented for the limits 
of the corresponding field strength regions including the transition points 
and for some value of the intermediate field strength in between. 
For each separate configuration the effect of the increasing field strength
consists in compressing the electronic distribution towards the $z$ axis. 
For the $1s2p_{-1}3d_{-2}$ configuration 
for which all single-electron binding energies increase unlimited 
for $\gamma\rightarrow\infty$ a 
shrinking process of this distribution in $z$ direction is also visible. 
For the $1s^22p_{-1}$ configuration this effect is not distinct for the
relevant field strengths.
For the $1s^22s$ state the opposite effect can be observed:
the $2s$ electronic charge distribution along the $z$ axis
expands slightly in weak magnetic fields.
A characteristic feature of the transition points is 
an inflation of the electronic distribution in $\rho$ direction 
during transitions from lower- to higher-field ground state configurations. 
This effect occurs due to the prevailing of the lowering in energy with
changing quantum numbers 
($m=0$ to $m=-1$ for the transition point $\gamma =0.17633$ and 
$S_z={\sum_{\mu=1}^3s_{z\mu}}=-1/2$ to $S_z=-3/2$ for $\gamma=2.153$) 
over the raising of the energy due to more extended charge 
distributions in the
$\rho$ direction. 

The total binding energies of the configurations 
$1s^22s$, $1s^2 2p_{-1}$,
$1s\downarrow 2s\downarrow 2p_{-1}\downarrow$, 
$1s\downarrow 2p_0\downarrow 2p_{-1}\downarrow$ and
$1s\downarrow 2p_{-1}\downarrow 3d_{-2}\downarrow$
are presented in figure 4. These values do not include spin
polarization terms and it can clearly be seen that the atomic ground
state in a magnetic field does in general not possess 
the largest binding energy.

Along with the total energy of the Li atom ground state 
we have obtained its ionization energies $E_I$ dependent on $\gamma$. 
The total energy values of the ground state of the ion ${\rm Li^+}$ 
are required for these  calculations. 
The set of the ground state configurations of this two-electron 
ion is analogous to those of the helium atom \cite{TKBHRW,Ivanov94} 
and consists of the zero-field ground state $1s^2$ and the 
strong field fully spin-polarized state $1s\downarrow2p_{-1}\downarrow$. 
Results of our calculations for these states are presented 
in table \ref{tab:lipgr}. 
The change of the ground state configuration takes place at 
$\gamma=2.071814$.
Comparing tables \ref{tab:ebgamma} and \ref{tab:lipgr} one 
obtains the dependence of the ionization energy of the ground
state of the Li atom on the magnetic field strength, as shown in figure 5.
This curve exhibits three distinct points marked by dotted vertical lines. 
The first of them (from left to right) corresponds to the change of the 
ground state configuration of the lithium atom from $1s^22s$ to $1s^22p_{-1}$. 
The second corresponds to the change of the $\rm Li^+$ ground 
state configuration from $1s^2$ to $1s\downarrow2p_{-1}$. 
And the third, very near to the second one, corresponds to 
the second change of the Li atom ground state configuration 
from $1s^22p_{-1}$ to $1s\downarrow 2p_{-1}\downarrow 3d_{-2}\downarrow$.
Table II provides the numerical data for the ionization energies.
Tables \ref{tab:ebgamma} and \ref{tab:lipgr} allow also obtaining 
ionization energies for other states presented in 
table \ref{tab:ebgamma}. 

In addition we show in figure 6 the total quadrupole moment 
\begin{eqnarray}
Q_{zz}={\langle}\Psi|3z^2-r^2|\Psi{\rangle},\ \ 
\ \ \ \ \ \ \ \ \ \ \ \ \ \ \ \ \ r^2=\rho^2+z^2
\label{eq:Qdef}
\end{eqnarray}
of different states of the atom as a function of the field strength.
These dependencies illustrate the changes in the density distribution of the 
electrons with increasing magnetic field strength. 
For weak and also to some extent for intermediate field strengths 
the main effect 
consists in compressing the wave function towards the $z$ axis. 
This results in increasing $Q_{zz}$ values and 
a sign change of $Q_{zz}$ for the states with initially negative 
$Q_{zz}$.
For $\gamma>10$ the continuing compression towards the $z$ axis 
practically does not affect $Q_{zz}$ due to the small values of 
${\langle}\rho^2{\rangle}$.
The values of $Q_{zz}$ decrease in this region 
for all the states considered with exception of the state 
$1s\downarrow 2p_0\downarrow 2p_{-1}\downarrow$.
This decrease of $Q_{zz}$ is associated with the decreasing 
value of ${\langle}z^2{\rangle}$
due to an increasing one-particle binding energy. 
For the states 
$1s^2 2p_{-1}$ and 
$1s\downarrow2p_{-1}\downarrow3d_{-2}\downarrow$ 
all these binding energies become infinite for infinite strong fields. 
This results in 
${Q_{zz}\rightarrow 0}$ as ${\gamma \rightarrow \infty}$. 
For the other states presented in Figure 6 at least one of the 
single-electron energies remains finite as ${\gamma \rightarrow \infty}$ 
and, in result, $Q_{zz}$ has a finite limit 
as ${\gamma \rightarrow \infty}$. 

\section{Summary and conclusions}

We have applied our 2D mesh Hartree-Fock method to a magnetized Li atom. 
The method is flexible enough to yield precise results for arbitrary 
field strengths and our calculations for the ground and several excited 
states are performed for magnetic field strengths ranging from zero 
up to  $2.3505\cdot 10^8$T ($\gamma=1000$). 
Our consideration was focused on the ground state of the Li atom. 
With increasing field strength this state undergoes two transitions 
involving three different electronic configurations. 
For weak fields up to $\gamma=0.17633$ the ground state arises from 
the field-free $1s^22s$ configuration. 
For intermediate fields ($0.17633<\gamma<2.1530$) the ground state is
constituted by the $1s^22p_{-1}$ configuration 
and for $\gamma>2.1530$ the ground state configuration is the fully 
spin-polarized $1s2p_{-1}3d_{-2}$ configuration. 
We provide arguments which show that 
this configuration must correspond to the ground state 
in the strong field limit. 
Generalizing these arguments we could derive the high-field ground state configuration 
of arbitrary fully spin polarized atoms which are constituted by certain tightly bound
hydrogen-like states.
For example for atoms with six electrons (i.e. C and C-like ions) 
the high-field ground state is given by the fully spin polarized 
$1s\downarrow2p_{-1}\downarrow3d_{-2}\downarrow4f_{-3}\downarrow5g_{-4}
\downarrow6h_{-5}\downarrow$ configuration.

We have also calculated HF energies for the two ${\rm Li^+}$ 
ground state configurations $1s^2$ and $1s\downarrow2p_{-1}\downarrow$. 
The first of them forms the ground state at $0\leq\gamma<2.071814$, 
the second one is the high-field ground state configuration 
for $\gamma>2.071814$. 
These calculations allowed us to obtain the Li atom ground state 
ionization energy $E_I$ dependent on the magnetic field strength. 
This dependence, opposite to the analogous dependence for the total and binding
energies is not monotoneous and contains both areas of 
increasing values of $E_I$ and 
a domain of decreasing behaviour between 
$\gamma=2.071814$ and $\gamma=2.1530$.
Furthermore we have studied the quadrupole moment of the atom and 
show how its complicated behaviour with changing field strength can be 
explained through the field dependence of the different HF orbitals.

Two remarks are in order. Our HF results do not include
the effects of correlation. To take into account the latter
would require a multiconfigurational approach which goes beyond
the scope of the present paper. We, however, do not expect that
the correlation energy changes our main conclusions like, for example, the
transitions in the ground states configurations or the behaviour of
the ionization energies depending on the field strength.
With increasing field strength the effective one particle
picture should be an increasingly better description of the wave function
and the percentage of the correlation energy should therefore
decrease.  For the case of hydrogen it is
well-known that in the high field regime ($\gamma >> 10^2$)
mass correction terms due to the finite nuclear mass become relevant
i.e. are no more negligible in
comparison with the Coulomb binding energies. The most important mass
corrections can be included by replacing the electron mass through its reduced
mass and results from the infinite nuclear mass calculations are related
to those with the reduced mass via a scaling relation. In the case of the 
much heavier Li atom these effects are expected to be much smaller.

Apart from the Li atom other species i.e. three-electron objects 
are expected to be in particular of astrophysical interest:
the three-electron ions formed by the nuclei He, C, O, and Ne 
possess a high abundance in the universe. 
To study these systems is the subject of a separate
investigation.

\vspace*{0.5cm}

\begin{center}
{\bf{Acknowledgment}}
\end{center}
One of the authors (M.V.I.) gratefully acknowledges financial support from the
Deutsche Forschungsgemeinschaft.

{}

\vspace*{2.0cm}

{\bf Figure Captions}

{\bf Figure 1.} Total energies of the Li atom as a function
of the magnetic field strength (solid lines marked by centered symbols). 
Dotted lines are energies of two electronic configurations of the 
$\rm Li^+$ ion: (a) low-field ground state $1s^2$; 
(b) high-field ground state $1s2p_{-1}$.

{\bf Figure 2.} The same as in figure 1 in the
relevant regime of transitions of the ground state configurations.

{\bf Figure 3.} Contour plots of the total electronic
densities for the ground state of the Li atom. 
The densities for neighbouring lines are different by a factor of $e$.

{\bf Figure 4.} Binding energies of various states of the Li atom 
as a function of the magnetic field strength. 

{\bf Figure 5.} Li atom ground state ionization energy $E_I$  
for a broad range of field strengths.
Transition points are marked by broken vertical lines. 
The first transition (from left to right) corresponds to the change of the 
ground state configuration from $1s^22s$ to $1s^22p_{-1}$. 
The second transition corresponds to the change of the
$\rm Li^+$ ground state configuration from $1s^2$ to $1s2p_{-1}$. 
Third transition of the Li ground state configuration 
from $1s^22p_{-1}$ to $1s2p_{-1}3d_{-2}$.

{\bf Figure 6.} Quadrupole moment of the Li atom depending on the 
magnetic field strength.


\newpage
\begin{table}
\caption{Total energies of several electronic ground and excited states
of the Li atom in the regime of field strength $\gamma =0,...,1000$}
\begin{tabular}{@{}llllllllll}
&\multicolumn{2}{c}{$1s^22s$}&\multicolumn{2}{c}{$1s^22p_{-1}$}
&\multicolumn{2}{c}{$1s2s2p_{-1}$}
&\multicolumn{2}{c}{$1s2p_02p_{-1}$}&$1s2p_{-1}3d_{-2}$\\
\cline{2-3}\cline{4-5}\cline{6-7}\cline{8-9}\cline{10-10}
$\gamma$
&$E$&$E$\cite{JonesOrtiz}
&$E$&$E$\cite{JonesOrtiz}
&$E$&$E$\cite{JonesOrtiz}
&$E$&$E$\cite{JonesOrtiz}
&$E$
\\
\noalign{\hrule}
0.0000
&  -7.43275
&   -7.4327
&  -7.36509
&   -7.3651
&  -5.35888
&   -5.3583
&  -5.23186
&   -5.2318
&  -5.08379
\\
0.0010
&  -7.43326
&
&  -7.36609
&
&  -5.36088
&
&  -5.23386
&
&  -5.08679
\\
0.0018
&  -7.43365
&   -7.4337
&  -7.36689
&   -7.3669
&  -5.36247
&   -5.3625
&  -5.23546
&   -5.2355
&  -5.08915
\\
0.0020
&  -7.43375
&
&  -7.36709
&
&  -5.36288
&
&  -5.23586
&
&  -5.08976
\\
0.0050
&  -7.43522
&
&  -7.37002
&
&  -5.36884
&
&  -5.24182
&
&  -5.09852
\\
0.0090
&  -7.43713
&   -7.4371
&  -7.37387
&   -7.3738
&  -5.37673
&   -5.3767
&  -5.24973
&   -5.2497
&  -5.10988
\\
0.0100
&  -7.43760
&
&  -7.37481
&
&  -5.37871
&
&  -5.25170
&
&  -5.11268
\\
0.0180
&  -7.44125
&   -7.4412
&  -7.38218
&   -7.3832
&  -5.39429
&   -5.3943
&  -5.26734
&   -5.2673
&  -5.13433
\\
0.0200
&  -7.44214
&
&  -7.38397
&
&  -5.39817
&
&  -5.27121
&
&  -5.13960
\\
0.0500
&  -7.45398
&
&  -7.40844
&
&  -5.45442
&
&  -5.32786
&
&  -5.21281
\\
0.0540
&  -7.45537
&   -7.4553
&  -7.41141
&   -7.4114
&  -5.46168
&   -5.4617
&  -5.33521
&   -5.3352
&  -5.22199
\\
0.1000
&  -7.46857
&
&  -7.44176
&
&  -5.54149
&
&  -5.41643
&
&  -5.32140
\\
0.1260
&  -7.47408
&   -7.4739
&  -7.45650
&   -7.4565
&  -5.58376
&   -5.5837
&  -5.45992
&   -5.4599
&  -5.37371
\\
0.17633
&  -7.48162
&
&  -7.48162
&
&
&
&
&
&
\\
0.1800
&  -7.48204
&   -7.4814
&  -7.48330
&   -7.4832
&  -5.66585
&   -5.6656
&  -5.54555
&   -5.5455
&  -5.47568
\\
0.2000
&  -7.48400
&
&  -7.49220
&
&  -5.69451
&
&  -5.57585
&
&  -5.51151
\\
0.5000
&  -7.47741
&
&  -7.58790
&
&  -6.04787
&
&  -5.96957
&
&  -5.97052
\\
0.5400
&  -7.47351
&   -7.4731
&  -7.59709
&   -7.5965
&  -6.08746
&   -6.0844
&  -6.01603
&   -6.0159
&  -6.02414
\\
0.9000
&  -7.42504
&   -7.4240
&  -7.65628
&   -7.6563
&  -6.40175
&   -6.3993
&  -6.39613
&   -6.3956
&  -6.46061
\\
1.0000
&  -7.40879
&
&  -7.66653
&
&  -6.48029
&
&  -6.49248
&
&  -6.57081
\\
1.2600
&  -7.36226
&   -7.3609
&  -7.68288
&   -7.6820
&  -6.67494
&   -6.6720
&  -6.72931
&   -6.7284
&  -6.84122
\\
1.8000
&  -7.24603
&   -7.2446
&  -7.67657
&   -7.6747
&  -7.05430
&   -7.0403
&  -7.17326
&   -7.1711
&  -7.34723
\\
2.0000
&  -7.19621
&
&  -7.66246
&
&  -7.18889
&
&  -7.32494
&
&  -7.52003
\\
2.071814
&  -7.17745
&
&  -7.65600
&
&  -7.23650
&
&  -7.37799
&
&  -7.58047
\\
2.1530
&
&
&  -7.64785
&
&
&
&
&
&  -7.64785
\\
2.1600
&
&
&  -7.64711
&   -7.6459
&  -7.29445
&   -7.2826
&  -7.44218
&   -7.4404
&  -7.65361
\\
2.5000
&  -7.05619
&
&  -7.60351
&
&  -7.51255
&
&  -7.71826
&
&  -7.92532
\\
3.0000
&  -6.89559
&
&  -7.51516
&
&  -7.81834
&
&  -8.00837
&
&  -8.29920
\\
3.6000
&  -6.67874
&   -6.6640
&  -7.37638
&   -7.3627
&  -8.16336
&   -8.1159
&  -8.37214
&   -8.3564
&  -8.71464
\\
3.9600
&
&
&  -7.27826
&   -7.2722
&  -8.35994
&   -8.3165
&  -8.57739
&   -8.5578
&  -8.94929
\\
4.3200
&
&
&  -7.17026
&   -7.1655
&  -8.54941
&   -8.5075
&  -8.77415
&   -8.7526
&  -9.17442
\\
4.6800
&
&
&  -7.05326
&   -7.0391
&  -8.73233
&   -8.6767
&  -8.96327
&   -8.9371
&  -9.39099
\\
5.0000
&  -6.08811
&
&  -6.94230
&
&  -8.88981
&
&  -9.12554
&
&  -9.57694
\\
5.0400
&
&
&  -6.92800
&   -6.9050
&  -8.90918
&   -8.8375
&  -9.14546
&   -9.1160
&  -9.59977
\\
5.4000
&  -5.90113
&   -5.8772
&  -6.79517
&   -6.7747
&  -9.08045
&   -9.0035
&  -9.32134
&   -9.2755
&  -9.80147
\\
7.0000
&  -5.08909
&
&  -6.12670
&
&  -9.78357
&
& -10.03896
&
& -10.62578
\\
   10.
&  -3.35777
&
&  -4.61777
&
& -10.91059
&
& -11.17886
&
& -11.93902
\\
   20.
&   3.49120
&
&   1.70565
&
& -13.69420
&
& -13.96582
&
& -15.16260
\\
   50.
&   27.6916
&
&  24.97942
&
&  -18.8012
&
&  -19.0436
&
&  -21.0505
\\
  100.
&    71.807
&
&   68.1735
&
&   -23.987
&
&  -24.1946
&
&  -27.0192
\\
  200.
&   164.371
&
&  159.5749
&
&   -30.559
&
&  -30.7327
&
&  -34.5850
\\
  500.
&    451.69
&
&  444.9033
&
&   -41.821
&
&   -41.959
&
&  -47.5583
\\
 1000.
&    939.54
&
& 930.84308
&
&    -52.65
&
&   -52.771
&
&  -60.0589
\end{tabular}
\label{tab:ebgamma}
\end{table}

\newpage
\begin{table}
\caption{Energies of the low- and high-field ground states of 
the ion ${\rm Li^+}$ and the ionization energy of the ground state of the Li atom 
$E_I$ for field strengths $\gamma = 0,...,1000$.}
\begin{tabular}{@{}llllllllll}
$\gamma$&$1s^2$&$1s2p_{-1}$&$E_I({\rm Li})$
\\
\noalign{\hrule}
0.0000
&  -7.23642
&  -5.02469
&   0.19633
\\
0.0010
&  -7.23642
&  -5.02619
&   0.19684
\\
0.0020
&  -7.23642
&  -5.02769
&   0.19733
\\
0.0050
&  -7.23641
&  -5.03218
&   0.19881
\\
0.0100
&  -7.23641
&  -5.03963
&   0.20119
\\
0.0200
&  -7.23639
&  -5.05442
&   0.20575
\\
0.0500
&  -7.23623
&  -5.09797
&   0.21775
\\
0.1000
&  -7.23567
&  -5.16789
&   0.23290
\\
0.17633
&  -7.23411
&  -5.26874
&   0.24751
\\
0.2000
&  -7.23345
&  -5.29873
&   0.25875
\\
0.5000
&  -7.21798
&  -5.64006
&   0.36992
\\
1.0000
&  -7.16401
&  -6.11462
&   0.50252
\\
2.0000
&  -6.96300
&  -6.89408
&   0.69946
\\
2.071814
&  -6.94440
&  -6.94440
&   0.71160
\\
2.1530
&  -6.92278
&  -7.00057
&   0.64729
\\
2.5000
&  -6.82347
&  -7.23258
&   0.69275
\\
3.0000
&  -6.66237
&  -7.54672
&   0.75248
\\
5.0000
&  -5.85051
&  -8.62943
&   0.94751
\\
7.0000
&  -4.84725
&  -9.52492
&   1.10086
\\
   10.
&  -3.11092
& -10.65131
&   1.28771
\\
   20.
&   3.74896
& -13.42974
&   1.73286
\\
   50.
&  27.96465
& -18.52548
&   2.5250
\\
  100.
&  72.09337
& -23.69994
&   3.3193
\\
  200.
& 164.66867
& -30.26077
&   4.3242
\\
  500.
&  452.0032
& -41.50393
&   6.0544
\\
 1000.
& 939.87976
&  -52.3230
&   7.7359
\end{tabular}
\label{tab:lipgr}
\end{table}

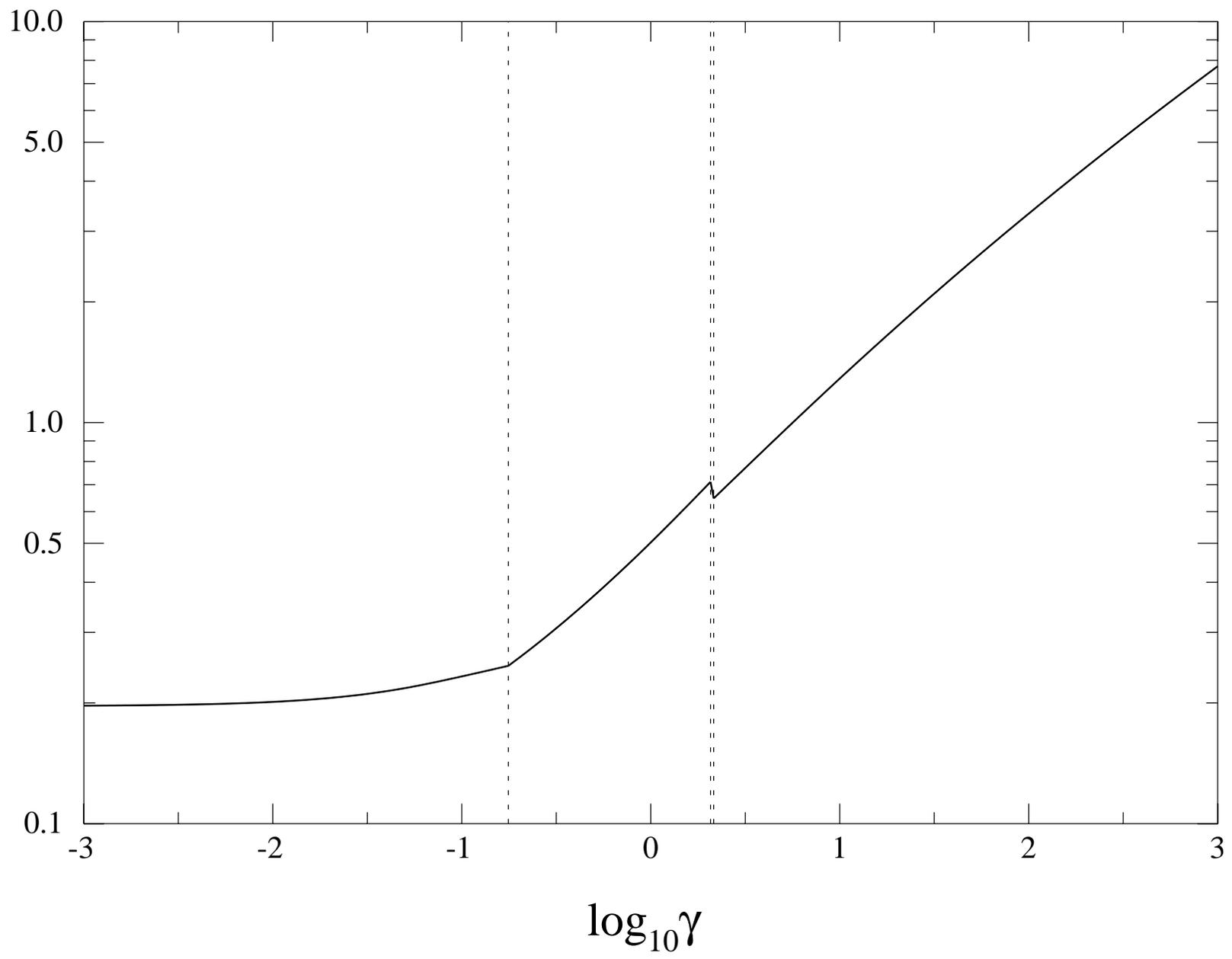
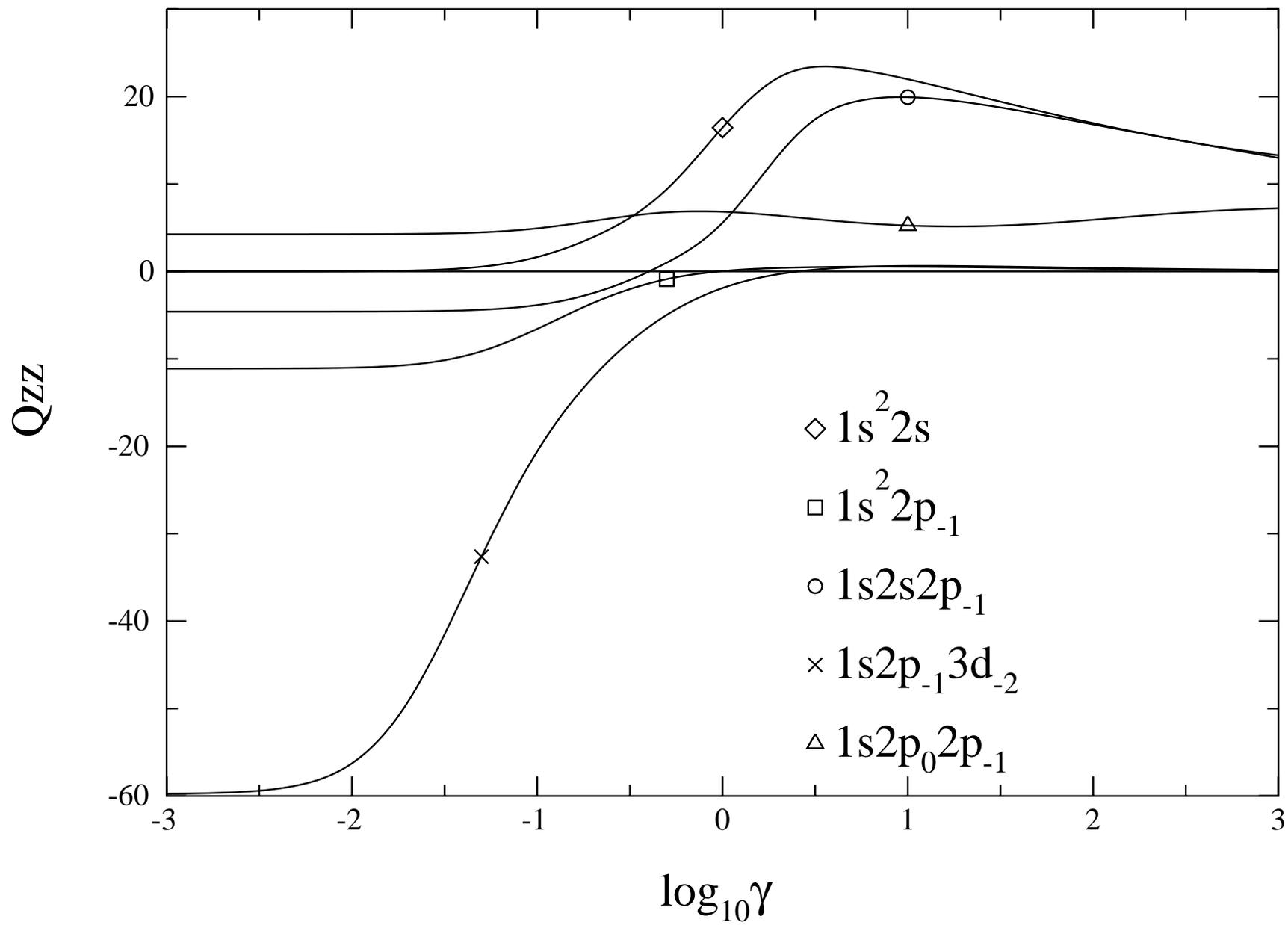
\end{document}